\documentclass[letter]{aa}

\usepackage{graphicx}
\usepackage{txfonts}
\usepackage{color}

\begin{document}

\title{An interesting case of the formation and evolution\\ of a barred galaxy in the cosmological context
}

\author{Ewa L. {\L}okas
}

\institute{Nicolaus Copernicus Astronomical Center, Polish Academy of Sciences,
Bartycka 18, 00-716 Warsaw, Poland\\
\email{lokas@camk.edu.pl}}


\abstract{
Elongated, bar-like galaxies without a significant disk component, with little rotation support and no gas, often form
as a result of tidal interactions with a galaxy cluster, as was recently demonstrated using the IllustrisTNG-100
simulation. Galaxies that exhibit similar properties are, however, also found to be infalling into the cluster
for the first time. We use the same simulation to study in detail the history of such a galaxy over cosmic
time in order to determine its origin. The bar appears to be triggered at $t=6.8$ Gyr by the combined effect of the
last significant merger with a subhalo and the first passage of another dwarf satellite, both ten times less massive
than the galaxy. The satellites deposit all their gas in the galaxy, contributing to its third and last star-formation
episode, which perturbs the disk and may also contribute to the formation of the bar. The galaxy then starts to lose
its gas and dark matter due to its passage near a group of more massive galaxies. The strongest interaction involves a
galaxy 22 times more massive, leaving the barred galaxy with no gas and half of its maximum dark matter mass. During
this time, the bar grows steadily, seemingly unaffected by the interactions, although they may have aided its growth by
stripping the gas. The studied galaxy, together with two other similar objects briefly discussed in this
letter, suggest the existence of a new class of early-type barred galaxies and thereby demonstrate the
importance of interactions in galaxy formation and evolution.}


\maketitle

\section{Introduction}

A considerable number of galaxies in the Universe possess bars \citep{Buta2015}. These elongated structures are formed
in at least two ways: through the inherent instability of disks that are cold enough \citep{Hohl1971, Ostriker1973} or
via interactions with other objects \citep{Gerin1990, Noguchi1987, Noguchi1996, Miwa1998, Berentzen2004, Lang2014,
Lokas2014, Lokas2018}. Such environmental effects may include fly-bys of smaller satellites or bigger galaxies as well
as mergers. Another scenario shown to be very effective in inducing bars is the evolution of galaxies in clusters.
\citet{Lokas2016} used controlled simulations of Milky Way-like galaxies orbiting a Virgo-like cluster to demonstrate
that the bar can efficiently be produced during the first pericenter passage if the orbit is both tight enough and
prograde, even if the galaxy is resilient to bar formation in isolation.

This scenario was recently tested in the cosmological context by \citet{Lokas2020b} using IllustrisTNG simulations
\citep{Springel2018, Marinacci2018, Naiman2018, Nelson2018, Pillepich2018a}. In that work, we followed the evolution of
well-resolved galaxies in the most massive cluster of IllustrisTNG-100 and found that the tidal forces from the cluster
have a strong effect on the morphology and kinematics of the galaxies, transforming many of them from rotating disks to
prolate spheroids dominated by random motions. Among the most strongly evolved sample of galaxies with more than one
pericenter passage, we identified six tidally induced bars and six other barred galaxies that probably had their bars
enhanced by the tidal interaction with the cluster. We note that these bars were not included in recent
studies of barred galaxies with IllustrisTNG \citep{Rosas2020, Zhou2020, Zhao2020}, which focused on late-type objects.
Instead, the bars tidally induced by the cluster would be rather classified as early-type due to their predominantly
prolate shape, limited rotation, and lack of gas and star formation.

Unexpectedly, very similar barred galaxies can also be identified at present among the unevolved sample of cluster
galaxies that are infalling into the cluster for the first time and have not strongly interacted  with it. In this
letter, we look at one such object in order to study its history and determine what other scenarios could lead to the
formation of this interesting subclass of barred galaxies. The galaxy is referred to by its identification number in
the subhalo catalog of the last output of the IllustrisTNG-100 simulation, ID44, in a way that resembles the naming of
real galaxies in various catalogs. The names of other objects mentioned below will also contain their number in the
last simulation snapshot, unless noted otherwise.

\section{Evolution of the galaxy}

\begin{figure}
\centering
\includegraphics[width=6.9cm]{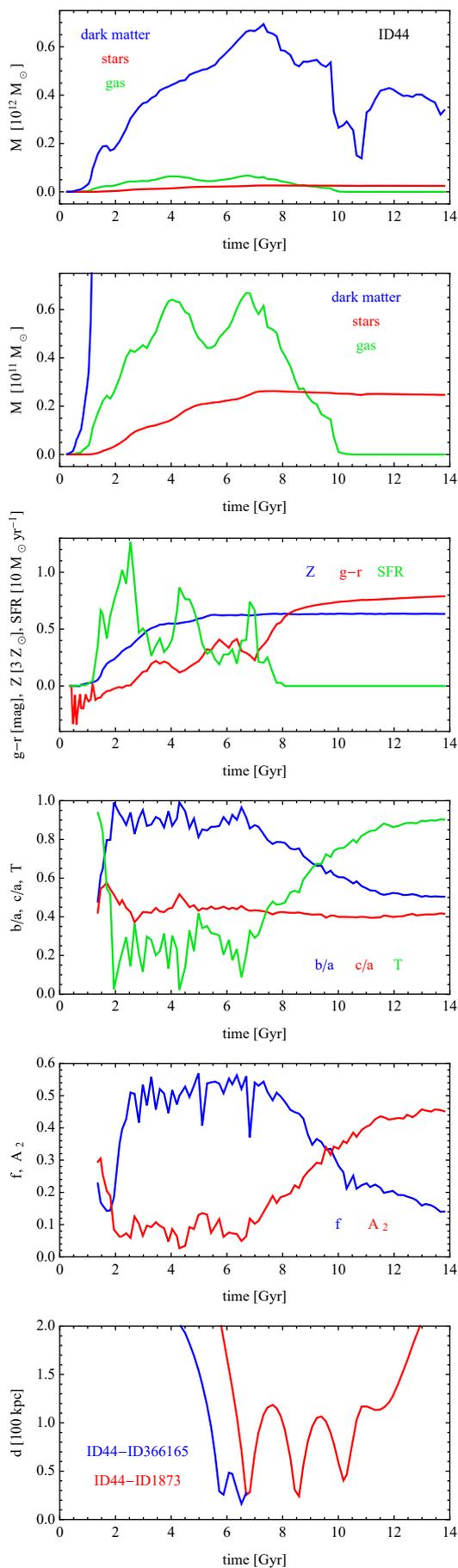}
\caption{Evolution of the galaxy ID44.}
\label{evolution}
\end{figure}

Figure~\ref{evolution} presents different properties of galaxy ID44 as a function of time obtained using the main
progenitor branch of the Sublink merger tree for this subhalo \citep{Rodriguez2015}. The two upper panels show the
behavior of the mass assigned to the galaxy divided into dark matter, stars, and gas, with two different ordinate
scales. The evolution of the dark mass (upper panel) shows two distinct phases, each lasting for about half of the
cosmic time. In the first phase (until about $t=7$ Gyr), the galaxy increases its dark mass and then, later on, it
mostly loses it. The gas content (second panel) also increases, albeit with some variation, until $t=7$ Gyr, and then
the gas is depleted until $t=10$ Gyr, when the galaxy does not possess any more gas. The stellar mass increases until
$t=7$ Gyr as well; later, it remains approximately constant, declining only slightly. The growth of the stellar mass is
not steady, rather, it accelerates in three epochs corresponding to three starbursts, as evidenced by the three peaks
of star formation rate (third panel). Star formation stops around $t=8$ Gyr. After that point, the galaxy becomes red
with its $(g-r)$ color increasing to 0.8 at the end of the evolution. The stellar metallicity remains roughly constant
from about $t=6$ Gyr. In this and the following panels, the measurements are carried out within two stellar half-mass
radii.

The next two panels of Fig.~\ref{evolution} describe the shape and kinematics of galaxy ID44 at different times using
parameters calculated as described in \citet{Genel2015} and provided together with the Illustris data release. For the
purpose of these measurements, the mass tensor of the stars was calculated and the galaxy was rotated to align it with
its principal axes. The eigenvalues of the mass tensor can then be used to estimate the axis ratios $c/a$ (shortest to
longest) and $b/a$ (intermediate to longest) and the triaxiality parameter, $T = [1-(b/a)^2]/[1-(c/a)^2]$. The
evolution of these three measures of shape is shown in the fourth panel of Fig.~\ref{evolution}. We can see that during
the evolution, the shape of the galaxy changes from oblate ($T<1/3$), to triaxial ($1/3 < T < 2/3$), and,
finally, prolate ($T>2/3$).

The strongly prolate shape at the end of the evolution suggests that within two stellar half-mass radii, the galaxy is
dominated by an elongated structure in the form of a bar. We confirm this finding by calculating the measure of the
strength of the bar as the $m=2$ mode of the Fourier decomposition of the surface density distribution of stellar
particles projected along the short axis: $A_2 (R) = | \Sigma_j m_j \exp(2 i \theta_j) |/\Sigma_j m_j$, where
$\theta_j$ is the azimuthal angle of the $j$th star, $m_j$ is its mass, and the sum goes up to the number of particles
in a given radial bin. The measurements within the cylindrical radius, $R,$ of two stellar half-mass radii are shown in
the fifth panel of Fig.~\ref{evolution}. Indeed, they are high at later times.

Since the formation of the bar should be accompanied by an increasing amount of radial motion of the stars, in order to
quantify the amount of rotation support in the galaxy, we used the rotation parameter, $f,$ defined as the fractional
mass of all stars with circularity parameter, $\epsilon > 0.7$. The evolution of this parameter is shown together with
$A_2$ in the fifth panel and we see that it decreases from rather high values of $f>0.5$ early on to $f<0.2$ at the end
of the evolution. We note that $f > 0.2$ is a reasonable threshold for defining late-type or disky galaxies in
Illustris \citep{Peschken2019, Peschken2020}, so $f < 0.2$ means that our evolved galaxy can no longer be considered a
disk.

\section{Properties of the bar}

\begin{figure*}
\centering
\includegraphics[width=5.6cm]{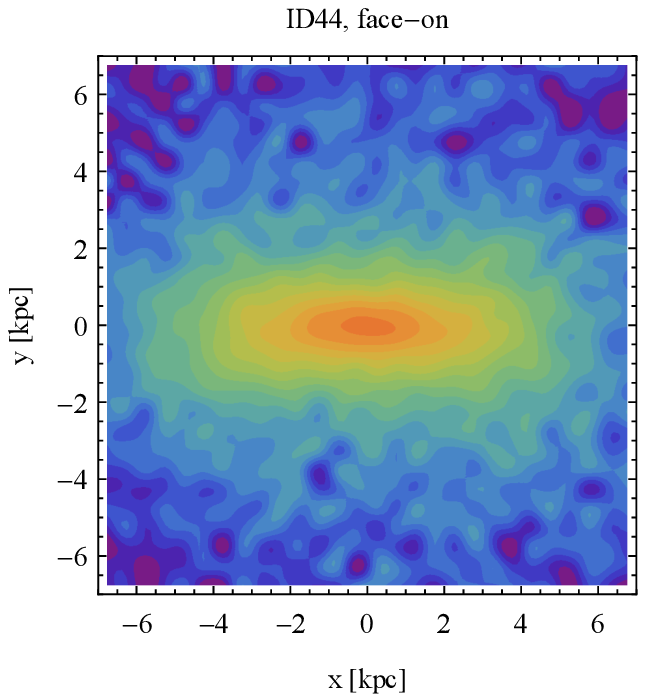}
\includegraphics[width=5.6cm]{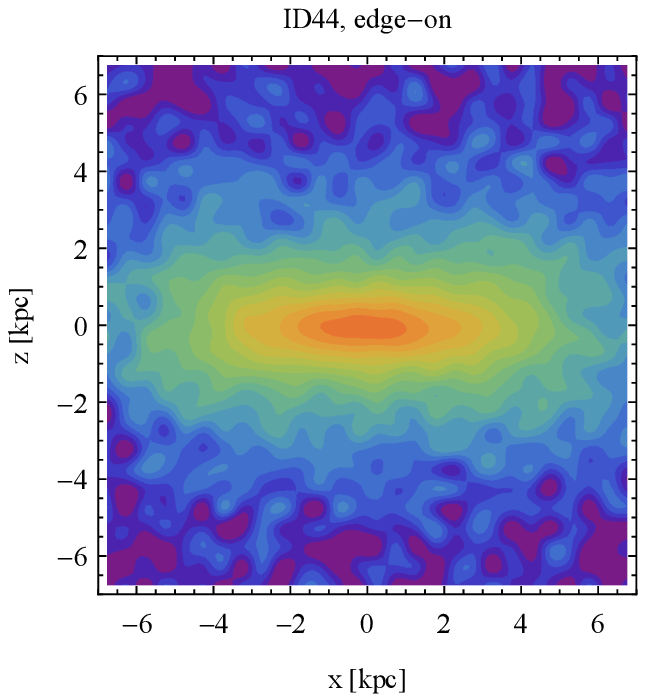}
\includegraphics[width=5.6cm]{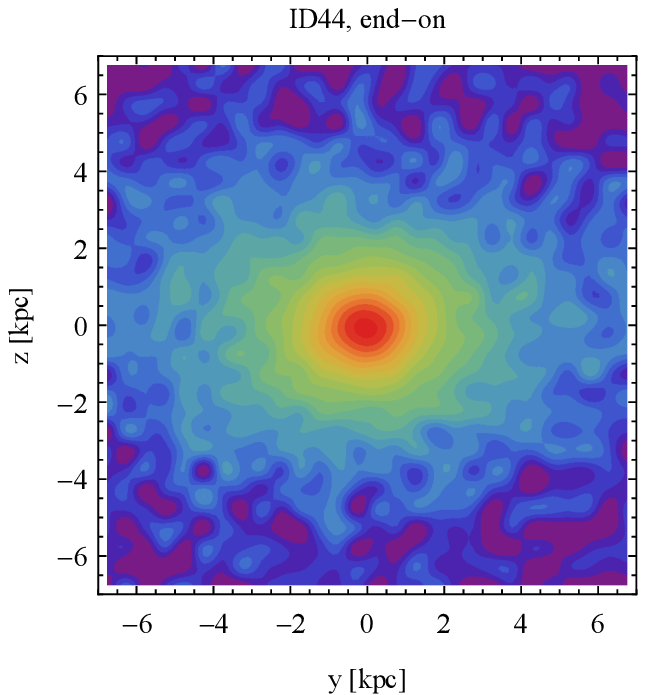}
\caption{Surface density distribution of the stellar component of ID44 viewed along the three principal axes at the
present time. Surface density, $\Sigma,$ is normalized to the central maximum value in the end-on view and the
contours are equally spaced in $\log \Sigma$.}
\label{surden}
\end{figure*}

In this section, we discuss the properties of the bar in greater detail. Figure~\ref{surden} shows the image of the
galaxy at the end of its evolution in terms of the surface density distribution of the stars in three projections,
along the shortest, intermediate, and the longest axis (from left to right). The image confirms that the galaxy  is
indeed a bar, in the sense that most of its stars contribute to the elongated shape, rather than a disk with an
embedded bar.

\begin{figure}
\centering
\includegraphics[width=7.2cm]{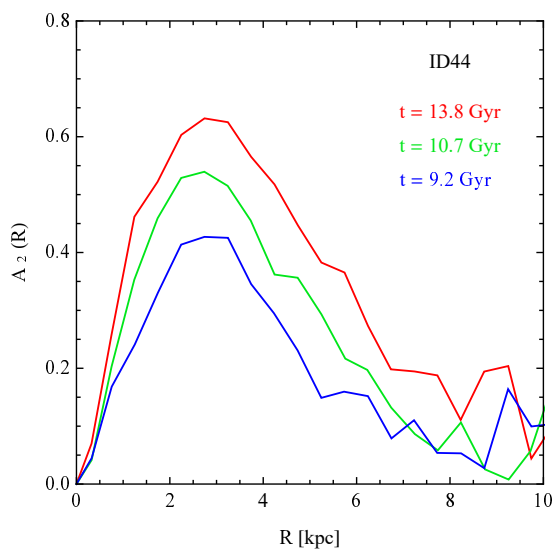}
\caption{Profiles of bar mode, $A_2 (R),$ for galaxy ID44 at different times. Measurements were carried out
in bins of $\Delta R = 0.5$ kpc.}
\label{a2profiles}
\end{figure}

The evolution of the bar can be described in detail by measuring the profiles of the bar mode, $A_2 (R),$ as a function
of the cylindrical radius, $R,$ rather than a single value in a wide radial bin shown in the fifth panel of
Fig.~\ref{evolution}. Examples of such profiles for ID44 at three different times are presented in
Fig.~\ref{a2profiles}. This figure demonstrates that $A_2(R)$ has a shape that is typical of bars, growing from zero at
small radii to a maximum and then decreasing again. As suggested by the results of the previous section, the strength
of the bar indeed grows in time since the values of the maximum of $A_2(R)$ are higher, and the bar length is larger,
at later times. The length of the bar can be estimated as the radius $R$ where $A_2(R)$ drops to half the maximum
value. For the three selected times shown in Fig.~\ref{a2profiles}, the bar length increases from about 4.5 to 6 kpc
and it is within the range of values measured for real galaxies \citep{Aguerri2005, Diaz2016, Font2017}. We note that
at the end of the evolution, the bar length is 2.7 times larger than the stellar half-mass radius at this time.

\begin{figure}
\centering
\hspace{0.7cm}
\includegraphics[width=3.5cm]{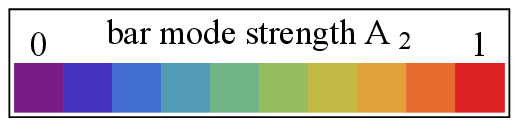}
\includegraphics[width=8.9cm]{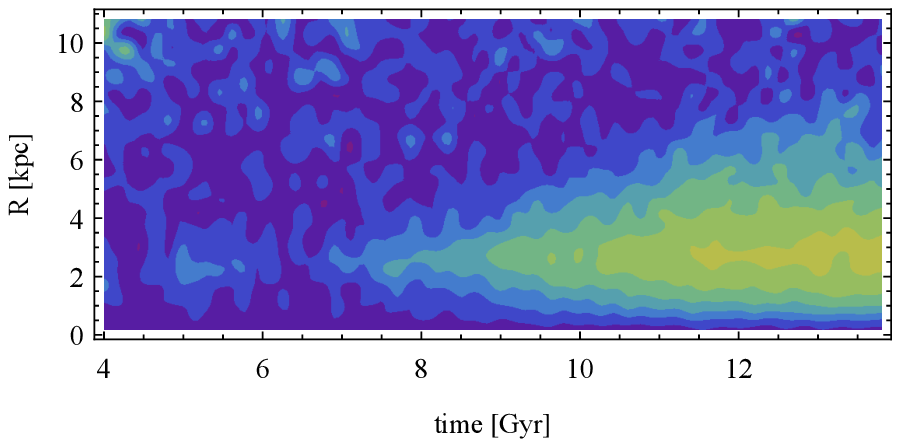}
\caption{Evolution of the profiles of the bar mode, $A_2 (R),$ of galaxy ID44 over time.}
\label{a2modestime}
\end{figure}

The dependence of bar strength and length on time can be fully appreciated by plotting the bar mode values as a
function both of radius and time. Such measurements are shown in the color-coded form in Fig.~\ref{a2modestime} for the
times between $t=4$ Gyr and $t=13.8$ Gyr. We can see that some elongation is already present around at $t=5$ Gyr but it
later disappears and the bar starts to continuously grow at about $t=6.8$ Gyr.

We also estimated the pattern speed $\Omega_{\rm b}$ of the bar. To this aim, for a first approximation, we used the
kinematic method proposed by \citet{Tremaine1984}. Such measurements are rather uncertain due to the low resolution of
the simulated galaxy and they depend on a number of parameters, such as the chosen binning and inclination. The method,
however, gives quite low values of the order of 6 km s$^{-1}$ kpc$^{-1}$, which means that the pattern speed can be
measured directly by the change of the position of the bar major axis between simulation outputs in spite of the fact
that they are spaced quite widely in time. This more direct method gives $\Omega_{\rm b}=9$ km s$^{-1}$ kpc$^{-1}$ at
the end of evolution, a value that is similar to what is estimated by \citet{Peschken2019} for the bars in Illustris
galaxies, but rather low in comparison with values from real barred galaxies \citep{Font2017}. We note however, that
around $t=9$ Gyr, when the bar is much weaker, the pattern speed is much higher, namely, $\Omega_{\rm b}=17$ km
s$^{-1}$ kpc$^{-1}$, and starts to steadily decrease around $t=10$ Gyr as the bar becomes stronger, which is in
agreement with other studies of simulated bars \citep{Athanassoula2003}.

\section{Interaction with environment}

The properties of the barred galaxy ID44 are very similar to the tidally induced bars identified by \citet{Lokas2020b}
in the most massive cluster of the IllustrisTNG-100 simulation. These bars were induced by strong tidal interactions
with the cluster over multiple pericenter passages. Six such objects were identified and a further six likely had their
bars enhanced by such interactions. Contrary to these objects, ID44 has never strongly interacted with the cluster. At
present, it is found at its outskirts, at a distance of 2.2 Mpc from the cluster center, comparable to the cluster
virial radius. Still, the bar could be induced tidally by interactions with other objects. In order to determine if
this is indeed the case, we studied the interactions of ID44 with other galaxies during its lifetime.

The last significant merger the galaxy experienced took place at $t=6.8$ Gyr. The satellite that merged with ID44 at
that time was last identified at one output step earlier (0.1 Gyr before), as ID366165. The last panel of
Fig.~\ref{evolution} displays its physical distance from ID44 as a function of time. Its total mass during infall,
when it was at $d=200$ kpc from ID44 (and therefore, when it was not yet heavily stripped by ID44) was $4.9 \times
10^{10}$ M$_\odot$ , which is equal to 9\% of the mass of ID44 at that time. As we can see from the last panel of
Fig.~\ref{evolution}, the satellite merged with ID44 after its second pericenter passage around it. At the same time,
ID44 was approached by another satellite, ID1873, of a very similar mass at $d=200$ kpc. This satellite does not merge
with ID44, but as shown in Fig.~\ref{evolution}, it completes three close pericenter passages around ID44, it
is later ejected toward the outskirts of ID44, and survives until the present. Both these satellites have probably
caused enough distortion in the disk of ID44 to seed the bar, as it started to grow around this time. The second
surviving satellite may have additionally contributed to the growth of the bar during its second pericenter passage,
although its mass was then much smaller. Both satellites deposited all their gas in ID44, which probably
contributed to igniting the third (that is, the last) of its star-formation bursts. The distribution of the densest gas
in ID44 at that time does not coincide with the center of the stellar distribution, so the new stars form off-center,
which may have also contributed to the formation of the bar.

The second important period of interactions in the lifetime of ID44 took place between 8 and 11 Gyr. During this time,
the galaxy passes close to a group of massive galaxies centered on a very massive elliptical (ID2). The interaction
causes ID44 to lose all its gas and a significant fraction of dark matter, as illustrated in the first two panels of
Fig.~\ref{evolution} depicting the mass evolution of the galaxy. The strongest interaction takes place at $t=10.8$ Gyr,
as indicated by the strongest dip in the dark matter mass assigned to ID44 at that time and even a small dent in the
stellar mass. At this moment, ID44 comes as close as about 100 kpc to ID8, which then had the total mass of $3.6 \times
10^{12}$ M$_\odot$, that is, 22 times greater than ID44. We note that the interactions with these more massive galaxies
appear to have little effect on the bar as it seems to grow steadily during this period. They may have, in fact, helped
the bar grow by stripping the gas since galaxies with smaller gas fractions are known to form bars more easily
\citep{Shlosman1993, Athanassoula2013, Lokas2020a}.

\section{Conclusions}

We analyzed the properties and evolution of galaxy ID44 in the IllustrisTNG-100 simulation. The galaxy was chosen as an
interesting example with properties that are similar to bars tidally induced by the most massive cluster in this
simulation and identified by \citet{Lokas2020b}. The stellar component of the galaxy is elongated and similar to a bar
but with hardly any disk component left. The galaxy is not strongly supported by rotation and contains no gas. In spite
of these similarities, the galaxy has never interacted with the cluster and is presently at its first infall, located
at its outskirts, at the physical distance of the cluster virial radius. We found that the bar was also tidally
induced, but by interactions with two satellites ten times smaller than ID44, one of which ended up merging with it.
The gas was later stripped during an interaction with a group of galaxies much bigger than ID44. However, the galaxy
cannot be considered as an example of group pre-processing, which is believed to be one of the mechanisms shaping the
galaxies in clusters \citep{Benavides2020}, as it never became a member of the group and has not been accreted onto the
cluster together with the other member galaxies; instead, it only passes within their vicinity.

ID44 is not the only case of galaxies that have not experienced even a single pericenter passage around the cluster but
nonetheless exhibit prolate shapes and limited rotation. There are a few similar objects in Fig. 9 of
\citet{Lokas2020b}, however, ID44 is the one with the highest triaxiality parameter $T=0.9$. Two further examples with
bar-like shapes and $T>0.8$ are ID55 and ID91. The former ends up beyond the virial radius of the cluster and seems to
have the bar induced by the first passage of a satellite at $t=5.5$ Gyr, and then  it becomes, itself, a satellite of a
much bigger galaxy ID1 which enhances its bar and strips all of its gas and most of its dark matter. The latter finds
itself closer to the center of the cluster at present. It has a weak bar induced by a passage of a small satellite
around $t=4$ Gyr, which is then enhanced by a merger with another satellite at $t=7.6$ Gyr. Around $t=11.5$ Gyr, it
passes close to a group of galaxies with masses similar to itself, and loses all the gas as well as a
significant fraction of dark matter.

Such objects with similar properties are present around other clusters or groups as well. There seems therefore to be a
subclass of bars hosted by early-type galaxies forming in the IllustrisTNG simulation that were not included in
previous studies of barred galaxies in Illustris and IllustrisTNG \citep{Peschken2019, Rosas2020, Zhou2020, Zhao2020}
since these studies focused on late-type objects and considered only bars embedded in strongly rotating, disky
galaxies. These early-type bars clearly demonstrate the importance of interactions in the processes of galaxy formation
and evolution, and, therefore, certainly deserve further study.

\begin{acknowledgements}
I am grateful to the referee for useful comments and the IllustrisTNG team for making their simulations
publicly available.
\end{acknowledgements}


\begin{thebibliography}{}

\bibitem[{Aguerri et al.}(2005)]{Aguerri2005} Aguerri, J. A. L., Elias-Rosa, N., Corsini, E. M., \& Munoz-Tunon, C.
        2005, A\&A, 434, 109
\bibitem[{Athanassoula}(2003)]{Athanassoula2003} Athanassoula, E. 2003, MNRAS, 341, 1179
\bibitem[{Athanassoula et al.}(2013)]{Athanassoula2013} Athanassoula, E., Machado, R. E. G., \& Rodionov, S. A. 2013,
        MNRAS, 429, 1949
\bibitem[{Benavides et al.}(2020)]{Benavides2020} Benavides, J. A., Sales, L. V., \& Abadi, M. G. 2020, MNRAS, in press,
        arXiv:2005.05344
\bibitem[{Berentzen et al.}(2004)]{Berentzen2004} Berentzen, I., Athanassoula, E., Heller, C. H., \& Fricke, K. J.
        2004, MNRAS, 347, 220
\bibitem[{Buta et al.}(2015)]{Buta2015} Buta, R. J., Sheth, K., Athanassoula, E., et al. 2015, ApJS, 217, 32
\bibitem[{Diaz et al.}(2016)]{Diaz2016} Diaz-Garcia, S., Salo, H., Laurikainen, E., \& Herrera-Endoqui, M.
        2016, A\&A, 587, A160
\bibitem[{Font et al.}(2017)]{Font2017} Font, J., Beckman, J. E., Martinez-Valpuesta, I., et al. 2017, ApJ, 835, 279
\bibitem[{Genel et al.}(2015)]{Genel2015} Genel, S., Fall, S. M., Hernquist, L., et al. 2015, ApJ, 804, L40
\bibitem[{Gerin et al.}(1990)]{Gerin1990} Gerin, M., Combes, F., \& Athanassoula, E. 1990, A\&A, 230, 37
\bibitem[{Hohl}(1971)]{Hohl1971} Hohl, F. 1971, ApJ, 168, 343
\bibitem[{Lang et al.}(2014)]{Lang2014} Lang, M., Holley-Bockelmann, K., \& Sinha, M. 2014, ApJ, 790, L33
\bibitem[{{\L}okas}(2018)]{Lokas2018} {\L}okas, E. L. 2018, ApJ, 857, 6
\bibitem[{{\L}okas}(2020a)]{Lokas2020a} {\L}okas, E. L. 2020a, A\&A, 634, A122
\bibitem[{{\L}okas}(2020b)]{Lokas2020b} {\L}okas, E. L. 2020b, A\&A, 638, A133
\bibitem[{{\L}okas et al.}(2014)]{Lokas2014} {\L}okas, E. L., Athanassoula, E., Debattista, V. P., et al. 2014,
        MNRAS, 445, 1339
\bibitem[{{\L}okas et al.}(2016)]{Lokas2016} {\L}okas, E. L., Ebrov\'{a}, I., del Pino, A., et al. 2016,
        ApJ, 826, 227
\bibitem[{Marinacci et al.}(2018)]{Marinacci2018} Marinacci, F., Vogelsberger, M., Pakmor, R., et al. 2018,
        MNRAS, 480, 5113
\bibitem[{Miwa \& Noguchi}(1998)]{Miwa1998} Miwa, T., \& Noguchi, M. 1998, ApJ, 499, 149
\bibitem[{Naiman et al.}(2018)]{Naiman2018} Naiman, J. P., Pillepich, A., Springel, V., et al., 2018, MNRAS, 477, 1206
\bibitem[{Nelson et al.}(2018)]{Nelson2018} Nelson, D., Pillepich, A., Springel, V., et al. 2018, MNRAS, 475, 624
\bibitem[{Noguchi}(1987)]{Noguchi1987} Noguchi, M. 1987, MNRAS, 228, 635
\bibitem[{Noguchi}(1996)]{Noguchi1996} Noguchi, M. 1996, ApJ, 469, 605
\bibitem[{Ostriker \& Peebles}(1973)]{Ostriker1973} Ostriker, J. P., \& Peebles, P. J. E. 1973, ApJ, 186, 467
\bibitem[{Peschken \& {\L}okas}(2019)]{Peschken2019} Peschken, N., \& {\L}okas, E. L. 2019, MNRAS, 483, 2721
\bibitem[{Peschken et al.}(2020)]{Peschken2020} Peschken, N., {\L}okas, E. L., \& Athanassoula, E. 2020, MNRAS,
        493, 1375
\bibitem[{Pillepich et al.}(2018a)]{Pillepich2018a} Pillepich, A., Nelson, D., Hernquist, L., et al. 2018a,
        MNRAS, 475, 648
\bibitem[{Rodriguez-Gomez et al.}(2015)]{Rodriguez2015} Rodriguez-Gomez, V., Genel, S., Vogelsberger, M., et al.
        2015, MNRAS, 449, 49
\bibitem[{Rosas-Guevara et al.}(2020)]{Rosas2020} Rosas-Guevara, Y., Bonoli, S., Dotti, M., et al. 2020, MNRAS, 491,
        2547
\bibitem[{Shlosman \& Noguchi }(1993)]{Shlosman1993} Shlosman, I., \& Noguchi, M. 1993, ApJ, 414, 474
\bibitem[{Springel et al.}(2018)]{Springel2018} Springel, V., Pakmor, R., Pillepich, A., et al. 2018, MNRAS, 475, 676
\bibitem[{Tremaine \& Weinberg}(1984)]{Tremaine1984} Tremaine, S., \& Weinberg, M. D. 1984, ApJ, 282, L5
\bibitem[{Zhao et al.}(2020)]{Zhao2020} Zhao, D., Du, M., Ho, L. C., Debattista, V. P., \& Shi, J. 2020,
        submitted to ApJ, arXiv:2009.06895
\bibitem[{Zhou et al.}(2020)]{Zhou2020} Zhou, Z.-B., Zhu, W., Wang, Y., \& Feng, L.-L. 2020, ApJ, 895, 92


\end{thebibliography}
\end{document}